\documentclass[12pt,preprint]{aastex}

\shorttitle{Astronomical Image Processing}
\shortauthors{Houde \& Vaillancourt}

\begin{document}

\title{Astronomical Image Processing with Array Detectors}

\author{Martin Houde}

\affil{The Department of Physics and Astronomy, The University of Western
Ontario, London, Ontario, Canada N6A 3K7}

\email{houde@astro.uwo.ca}

\author{John E. Vaillancourt}

\affil{Physics Department, California Institute of Technology, MS 320-47,
1200 E. California Blvd., Pasadena, CA 91125}

\email{johnv@submm.caltech.edu}

\begin{abstract}
We address the question of astronomical image processing from data
obtained with array detectors. We define and analyze the cases of
evenly, regularly, and irregularly sampled maps for idealized (i.e.,
infinite) and realistic (i.e., finite) detectors. We concentrate on
the effect of interpolation on the maps, and the choice of the kernel
used to accomplish this task. We show how the normalization intrinsic
to the interpolation process must be carefully accounted for when
dealing with irregularly sampled grids. We also analyze the effect
of missing or dead pixels in the array, and their consequences for
the Nyquist sampling criterion. 
\end{abstract}

\keywords{methods: data analysis --- techniques: image processing }

\section{Introduction}

The creation of smooth two-dimensional maps from a series of samples
measured at discrete points is a common problem in astronomical image
processing. The goal is to create a smooth map with the best possible
spatial resolution given a set of data sampled in two-dimensions.
The solution is complicated by the fact that the data are often not
sampled in a regular way even if the detector layout is regular. For
example, telescopes may be scanned or dithered to map areas larger
than the array, some instruments are unable to follow objects as they
rotate on the sky, or an array itself may contain flaws (i.e., missing
or dead pixels). The resulting two-dimensional sample pattern can
often appear quite irregular. 

Although the layout of most modern detector arrays (e.g., CCDs) can
reasonably be approximated as generating evenly sampled grids (ESG)
extending to infinity in all directions, observations with these arrays
will typically include a series of array translations and rotations.
Additionally, one may want to combine multiple images of the same
piece of sky in order to increase the signal-to-noise ratio. This
requires that the images be registered so that the same area of the
sky is being observed in each image. That is, any relative translation
and rotation of the array positions with respect to the sky must be
taken into account when combining the images. Unless the translation
and rotation operations are such that every pixel lies in a location
previously occupied by another pixel then the resulting sample pattern
is no longer an ESG.

The processing of data from any case other than an ESG requires performing
an interpolation. Some of these cases have been discussed by other
authors (e.g., \citet{Granrath 1998}). The interpolation (or smoothing)
necessarily has an effect on the spatial resolution of the resulting
map. For the case of any sampling pattern (ESG or otherwise) we wish
to address the following two questions: 1) How does one choose an
optimal kernel shape and size for the interpolation function? and
2) How does this kernel choice affect the spatial resolution of the
resulting map? We will concentrate on the case of a Gaussian kernel.

As mentioned above, the construction of maps from non-ESG sampled
data is generally done through an interpolation of the data using
a smoothing kernel (see \S \ref{sec:RSG} and \citet{Lombardi 2001}).
In this paper, we begin by reviewing the solution for ESGs using a
technique based on Fourier transforms (\S \ref{sec:ESG}) and extend
this technique in \S \ref{sec:RSG} to regularly spaced grids (RSGs),
which are composed of relatively translated ESGs. In \S \ref{sec:ISG}
we use the tools developed for studying RSGs and ESGs to analyze irregularly
sampled grids (ISGs) and explore the effects of missing samples in
a map. 

Throughout this paper we present examples that reference the bolometer
array used with SHARP. SHARP is a polarimeter module that is used
in conjunction with the SHARC-II camera, which is deployed at the
Caltech Submillimeter Observatory \citep{Dowell 2003}. For this,
the $12\times32$ SHARC-II detector array is optically split into
two $12\times12$ sub-arrays (and a section of $12\times8$ unused
pixels), which image two orthogonal linear polarization components
of radiation \citep{Novak 2004}. Although we concentrate on SHARP
maps, the results are applicable to any detector array or sampling
pattern.

\section{Mathematical Definitions}

Before embarking on the analysis of the evenly sampled grid (ESG)
we first introduce a set of definitions and functions that are central
to the development of the subsequent sections. Given a function $g\left(\mathbf{r}\right)$,
which is dependent on position $\mathbf{r}=x\mathbf{e}_{x}+y\mathbf{e}_{y}$,
($\mathbf{e}_{x}$ and $\mathbf{e}_{y}$ are the usual Cartesian unit
basis vectors), we define the Fourier transform pair%
\footnote{In this paper we use a lower case and the corresponding capital letter
for a function and its Fourier transform, respectively. %
} 

\begin{eqnarray}
g\left(\mathbf{r}\right) & = & \int_{-\infty}^{\infty}G\left(\mathbf{w}\right)e^{j2\pi\mathbf{w}\cdot\mathbf{r}}dudv\label{eq:g(r)}\\
G\left(\mathbf{w}\right) & = & \int_{-\infty}^{\infty}g\left(\mathbf{r}\right)e^{-j2\pi\mathbf{w}\cdot\mathbf{r}}dxdy,\label{eq:G(w)}\end{eqnarray}

\noindent with $\mathbf{w}=u\mathbf{e}_{x}+v\mathbf{e}_{y}$ the spatial
frequency vector. The digitization of a signal will invariably introduce
trains of Dirac distributions. For example, a two-dimensional Dirac
train of periods $l_{1}$ and $l_{2}$ along $\mathbf{e}_{x}$ and
$\mathbf{e}_{y}$, respectively, is defined such that

\begin{equation}
\sum_{i,k=-\infty}^{\infty}\delta\left(\mathbf{r}-\mathbf{r}_{ik}\right)\equiv\sum_{i,k=-\infty}^{\infty}\delta\left(x-il_{1}\right)\delta\left(y-kl_{2}\right),\label{eq:diractrain}\end{equation}

\noindent with the following Fourier transform relation (see Appendix)

\begin{equation}
\sum_{i,k=-\infty}^{\infty}\delta\left(\mathbf{r}-\mathbf{r}_{ik}\right)\Leftrightarrow\frac{1}{l_{1}l_{2}}\sum_{m,n=-\infty}^{\infty}\delta\left(\mathbf{w}-\mathbf{w}_{mn}\right),\label{eq:dtrain_ft}\end{equation}

\noindent where 

\begin{eqnarray}
\mathbf{r}_{ik} & = & il_{1}\mathbf{e}_{x}+kl_{2}\mathbf{e}_{y}\label{eq:r_ik}\\
\mathbf{w}_{mn} & = & \frac{m}{l_{1}}\mathbf{e}_{x}+\frac{n}{l_{2}}\mathbf{e}_{y}.\label{eq:w_mn}\end{eqnarray}

Another useful distribution is the flat-top window of length $\Delta l_{1}$
(in the $\mathbf{e}_{x}$ direction in this case), which we denote
by

\begin{equation}
\mathrm{rect\mathnormal{\left(\frac{x}{\Delta l_{1}}\right)=\left\{ \begin{array}{cc}
1, & \left|x\right|<\frac{\Delta l_{1}}{2}\\
0, & \left|x\right|>\frac{\Delta l_{1}}{2},\end{array}\right.}}\label{eq:rect}\end{equation}

\noindent and the corresponding Fourier transform pair

\begin{equation}
\mathrm{rect}\left(\frac{x}{\Delta l_{1}}\right)\Leftrightarrow\Delta l_{1}\,\mathrm{sinc}\mathnormal{\left(\pi u\Delta l_{1}\right)}\equiv\Delta l_{1}\frac{\sin\left(\pi u\Delta l_{1}\right)}{\pi u\Delta l_{1}}.\label{eq:rect_ft}\end{equation}

\section{The Evenly Sampled Grid}

\label{sec:ESG}The detection of a signal $s\left(\mathbf{r}\right)$
from an astronomical source is inevitably achieved through a series
of transformations. Mathematically speaking, the signal is first convolved
with the telescope transfer function $b\left(\mathbf{r}\right)$ such
that 

\begin{equation}
s^{\prime}\left(\mathbf{r}\right)=s\left(\mathbf{r}\right)\otimes b\left(\mathbf{r}\right),\label{eq:sprime}\end{equation}

\noindent where {}``$\otimes$'' stands for a convolution, while
the measured signal $t^{\prime}\left(\mathbf{r}\right)$ is a sampled,
pixel-integrated version of $s^{\prime}\left(\mathbf{r}\right)$.
For an ESG, the sampling is done in an even manner with a Dirac train
as defined in equations (\ref{eq:diractrain}) and (\ref{eq:dtrain_ft}).
More precisely, for rectangular pixels of widths $\Delta l_{1}$ and
$\Delta l_{2}$ we write

\begin{eqnarray}
t^{\prime}\left(\mathbf{r}\right) & = & t\left(\mathbf{r}\right)\cdot\sum_{i,k=-\infty}^{\infty}\delta\left(\mathbf{r}-\mathbf{r}_{ik}\right)\nonumber \\
 & = & \sum_{i,k=-\infty}^{\infty}t\left(\mathbf{r}_{ik}\right)\delta\left(\mathbf{r}-\mathbf{r}_{ik}\right),\label{eq:t'(r)_ESG}\end{eqnarray}

\noindent with

\begin{eqnarray}
t\left(\mathbf{r}\right) & = & s^{\prime}\left(\mathbf{r}\right)\otimes p\left(\mathbf{r}\right)\nonumber \\
 & = & \left[b\left(\mathbf{r}\right)\otimes p\left(\mathbf{r}\right)\right]\otimes s\left(\mathbf{r}\right)\label{eq:t(r)_conv}\\
p\left(\mathbf{r}\right) & = & \mathrm{rect\mathnormal{\left(\frac{x}{\Delta l_{1}}\right)}}\mathrm{rect}\mathnormal{\left(\frac{y}{\Delta l_{2}}\right).}\label{eq:p(r)}\end{eqnarray}

The convolution 

\begin{equation}
h\left(\mathbf{r}\right)\equiv\left[b\left(\mathbf{r}\right)\otimes p\left(\mathbf{r}\right)\right]\label{eq:psf}\end{equation}

\noindent stands for what is commonly described as the point spread
function (PSF). Using equations (\ref{eq:G(w)}) and (\ref{eq:dtrain_ft}),
and the properties of the Fourier transform for products and convolutions
of functions, we find that

\begin{eqnarray}
T^{\prime}\left(\mathbf{w}\right) & = & T\left(\mathbf{w}\right)\otimes\frac{1}{l_{1}l_{2}}\sum_{m,n=-\infty}^{\infty}\delta\left(\mathbf{w}-\mathbf{w}_{mn}\right)\nonumber \\
 & = & \frac{1}{l_{1}l_{2}}\sum_{m,n=-\infty}^{\infty}T\left(\mathbf{w}-\mathbf{w}_{mn}\right),\label{eq:T'(w)}\end{eqnarray}

\noindent with

\begin{eqnarray}
T\left(\mathbf{w}\right) & = & H\left(\mathbf{w}\right)S\left(\mathbf{w}\right)\label{eq:T(w)}\\
 & = & B\left(\mathbf{w}\right)P\left(\mathbf{w}\right)S\left(\mathbf{w}\right),\label{eq:BPS(w)}\end{eqnarray}

\noindent and

\begin{equation}
P\left(\mathbf{w}\right)=\Delta l_{1}\Delta l_{2}\,\mathrm{sinc\mathnormal{\left(\pi u\Delta l_{1}\right)}}\mathrm{sinc\mathnormal{\left(\pi v\Delta l_{2}\right).}}\label{eq:P(w)}\end{equation}

Equation (\ref{eq:t'(r)_ESG}) is only valid for the idealized case
of an infinite array. In reality this relation should be multiplied
by an aperture function of appropriate width and shape. Although we
will take this restriction into account when analyzing the effect
of missing pixels in \S \ref{sub:missing}, we will for the moment
simplify our analysis by assuming that the array is sufficiently large
so that equations (\ref{eq:t'(r)_ESG}) and (\ref{eq:T'(w)}) are
suitable approximations.

\subsection{Interpolation}

\label{sub:Interpolation}The ESG studied in the previous section
is the simplest representation that can be given for a sampled set
of data. As we will see in later sections, we will always seek to
transform more complicated forms of data grids (i.e., not evenly sampled
ones) into ESGs to facilitate analysis; this will invariably require
the interpolation of sampled quantities from different locations.
Also, one might inquire about quantities at positions where there
are no samples. For example, questions such as {}``What is the intensity
at position A, where there is no sample, and how does it compare to
the flux at position B, C, and D on this map?'' are common when analyzing
astronomical images. It is, therefore, often necessary to generate
a new interpolated map from the data set expressed through equation
(\ref{eq:t'(r)_ESG}). 

Given a weighting function $w\left(\mathbf{r}\right)$, any value
$z_{\mathrm{int}}\left(\mathbf{r}\right)$ to be assigned to an interpolated
point can be expressed as

\begin{equation}
z_{\mathrm{int}}\left(\mathbf{r}\right)=n\left(\mathbf{r}\right)\cdot\sum_{i=1}^{n}z\left(\mathbf{r}_{i}\right)w\left(\mathbf{r}-\mathbf{r}_{i}\right),\label{eq:z_int(r)}\end{equation}

\noindent where $z\left(\mathbf{r}_{i}\right)$ is the value associated
with the $i$th of the $n$ data points used for the interpolation.
The quantity

\begin{equation}
n\left(\mathbf{r}\right)=\left[\sum_{i=1}^{n}w\left(\mathbf{r}-\mathbf{r}_{i}\right)\right]^{-1}\label{eq:n(r)_ESG}\end{equation}

\noindent is the normalization factor, which is a function of the
position of interpolation. The generation of an interpolated map is
equivalent to the convolution of the initial data set with the weighting
function followed by the normalization and re-sampling of the data.
This can be ascertained through a comparison of equation (\ref{eq:z_int(r)})
with

\begin{eqnarray}
t_{\mathrm{int}}\left(\mathbf{r}\right) & = & \sum_{s,t=-\infty}^{\infty}\delta\left(\mathbf{r}-\mathbf{r}_{st}-\mathbf{a}_{pq}\right)\cdot\left\{ n\left(\mathbf{r}\right)\cdot\left[t^{\prime}\left(\mathbf{r}\right)\otimes w\left(\mathbf{r}\right)\right]\right\} \nonumber \\
 & = & \sum_{s,t=-\infty}^{\infty}\delta\left(\mathbf{r}-\mathbf{r}_{st}-\mathbf{a}_{pq}\right)\cdot\left(n\left(\mathbf{r}\right)\cdot\left\{ \left[t\left(\mathbf{r}\right)\cdot\sum_{i,k=-\infty}^{\infty}\delta\left(\mathbf{r}-\mathbf{r}_{ik}\right)\right]\otimes w\left(\mathbf{r}\right)\right\} \right)\nonumber \\
 & = & \sum_{s,t=-\infty}^{\infty}\delta\left(\mathbf{r}-\mathbf{r}_{st}-\mathbf{a}_{pq}\right)\cdot\left[n\left(\mathbf{r}\right)\cdot\sum_{i,k=-\infty}^{\infty}t\left(\mathbf{r}_{ik}\right)w\left(\mathbf{r}-\mathbf{r}_{ik}\right)\right],\label{eq:tint(r)}\end{eqnarray}

\noindent where $\mathbf{r}_{st}$ is defined as in equation (\ref{eq:r_ik})
and

\begin{equation}
\mathbf{a}_{pq}=\frac{l_{1}}{p}\mathbf{e}_{x}+\frac{l_{2}}{q}\mathbf{e}_{y}\label{eq:a_pq}\end{equation}

\noindent is the displacement vector specifying the position of the
interpolated grid $t_{\mathrm{int}}\left(\mathbf{r}\right)$ in the
relation to the initial grid. It is important to realize that because
of the evenness in the sampling distribution of the original map $t^{\prime}\left(\mathbf{r}\right)$
the normalization factor $n\left(\mathbf{r}\right)$ will be periodic
in character with the same periods (i.e., $l_{1}$ and $l_{2}$) as
the original sampling Dirac train%
\footnote{It should be noted that the normalization function will be constant
for an infinite grid when $W\left(\mathbf{w}\right)=0$ for $\left|u\right|>\left(2l_{1}\right)^{-1}$
or $\left|v\right|>\left(2l_{2}\right)^{-1}$ (e.g., \emph{sinc} weighting
functions of corresponding widths in normal space). This condition
must be strictly enforced in order to obtain a constant normalization
factor while satisfying the Nyquist sampling criterion.%
}. As a consequence, it will take a common value for all interpolated
points similarly located within a one-period segment anywhere on the
grid (see Appendix). More precisely, data resulting from interpolations
at points at $\mathbf{r}$ and $\mathbf{r}+\mathbf{r}_{ik}$, for
any integer $i$ and $k$ when $\mathbf{r}_{ik}$ is defined as in
equation (\ref{eq:r_ik}), will have the same normalization factor.
Therefore, when the re-sampling is done using the same spatial sampling
rate as for the original grid (as is the case in eq. {[}\ref{eq:tint(r)}])
we can write Fourier transform of $t_{\mathrm{int}}\left(\mathbf{r}\right)$
as

\begin{equation}
T_{\mathrm{int}}\left(\mathbf{w}\right)=\left[\frac{1}{l_{1}l_{2}}\sum_{s,t=-\infty}^{\infty}\delta\left(\mathbf{w}-\mathbf{w}_{st}\right)e^{-j2\pi\mathbf{w}_{st}\cdot\mathbf{a}_{pq}}\right]\otimes\left[W\left(\mathbf{w}\right)\cdot\frac{c}{l_{1}l_{2}}\sum_{m,n=-\infty}^{\infty}T\left(\mathbf{w}-\mathbf{w}_{mn}\right)\right],\label{eq:Tint(w)}\end{equation}

\noindent where $c$ is the constant value associated with $n\left(\mathbf{r}\right)$
for this particular re-sampling process. Equation (\ref{eq:Tint(w)})
contains multiple copies of $T\left(\mathbf{w}\right)$, one for each
pair of $m$ and $n$. If the Nyquist sampling criterion is satisfied
(see Appendix), then the high frequency copies may be removed with
negligible aliasing by choosing the weighting function such that $W\left(\mathbf{w}\right)\sim0$
when $\left|\mathbf{w}\right|>\left|\mathbf{w}_{mn}\right|/2$ (when
$m\neq0$ or $n\neq0$). Equation (\ref{eq:Tint(w)}) then simplifies
to

\begin{equation}
T_{\mathrm{int}}\left(\mathbf{w}\right)=\frac{c}{\left(l_{1}l_{2}\right)^{2}}\sum_{s,t=-\infty}^{\infty}W\left(\mathbf{w}-\mathbf{w}_{st}\right)T\left(\mathbf{w}-\mathbf{w}_{st}\right)e^{-j2\pi\mathbf{w}_{st}\cdot\mathbf{a}_{pq}},\label{eq:Tint(w)_app}\end{equation}

\noindent and

\begin{equation}
t_{\mathrm{int}}\left(\mathbf{r}\right)=\left[t\left(\mathbf{r}\right)\otimes w\left(\mathbf{r}\right)\right]\cdot\frac{c}{l_{1}l_{2}}\sum_{i,k=-\infty}^{\infty}\delta\left(\mathbf{r}-\mathbf{r}_{ik}-\mathbf{a}_{pq}\right).\label{eq:tint(r)_app}\end{equation}

The only difference between $t_{\mathrm{int}}\left(\mathbf{r}\right)$
and $t^{\prime}\left(\mathbf{r}\right)$ (see eq. {[}\ref{eq:t'(r)_ESG}]),
besides the overall scaling factor and translation, is the presence
of the convolution by $w\left(\mathbf{r}\right)$ for the former.
It is therefore apparent that $w\left(\mathbf{r}\right)$ can serve
not only as a weighting function for interpolation but also as a smoothing
kernel, as its effect is functionally similar to that of the PSF $h\left(\mathbf{r}\right)$
or any other function that can be applied to $t\left(\mathbf{r}\right)$
(see eq. {[}\ref{eq:t(r)_conv}]) before or during the sampling process
leading to $t^{\prime}\left(\mathbf{r}\right)$. It therefore follows
that the weighting functions also possesses spectral filtering qualities,
as the base spectrum $T\left(\mathbf{w}\right)$ is multiplied by
its Fourier transform $W\left(\mathbf{w}\right)$ (see eq. {[}\ref{eq:Tint(w)_app}])
. One can, in fact, take advantage of this property in some cases.
For example, the extraction of a signal from noise can be optimized
by matching the spectral shape of the Fourier transform of the weighting
function (more appropriately named the {}``filter'' in this case)
to that of the signal itself (if such information is available \emph{a
priori}). This is a result commonly established through the so-called
matched filter theorem \citep{Haykin 1983}. It is to be noted, however,
that optimization of the signal-to-noise ratio through filtering is
not our goal. As will be made evident in \S\ref{sub:weight}, besides
its fundamental role in the interpolation process we are also concerned
with determining the effects of the weighting function on the spatial
resolution of a map. In general, the spatial extent of the smoothing
kernel will always be significantly smaller than that of the optimized
matched filter. 

We now investigate the case of a map resulting from a re-sampling
process where we seek to increase the density of samples. For example,
a map with half the sampling periods as the original (i.e., of periods
$l_{1}/2$ and $l_{2}/2$) will consist of the combination of four
different re-sampled maps $t_{1}\left(\mathbf{r}\right)$, $t_{2}\left(\mathbf{r}\right)$,
$t_{3}\left(\mathbf{r}\right)$, and $t_{4}\left(\mathbf{r}\right)$
that all share the same sampling periods as the original map, but
translated relative to each other. More precisely, we define

\begin{eqnarray}
t_{1}\left(\mathbf{r}\right) & = & t_{\mathrm{int}}\left(\mathbf{r}\right)\left|_{p,q\rightarrow\infty}\right.\label{eq:t1(r)_ESG}\\
t_{2}\left(\mathbf{r}\right) & = & t_{\mathrm{int}}\left(\mathbf{r}\right)\left|_{p=2,q\rightarrow\infty}\right.\label{eq:t2(r)_ESG}\\
t_{3}\left(\mathbf{r}\right) & = & t_{\mathrm{int}}\left(\mathbf{r}\right)\left|_{p\rightarrow\infty,q=2}\right.\label{eq:t3(r)_ESG}\\
t_{4}\left(\mathbf{r}\right) & = & t_{\mathrm{int}}\left(\mathbf{r}\right)\left|_{p=2,q=2}\right..\label{eq:t4(r)_ESG}\end{eqnarray}

In other words, $t_{1}\left(\mathbf{r}\right)$ is re-sampled at the
same positions as the original map while $t_{2}\left(\mathbf{r}\right)$,
$t_{3}\left(\mathbf{r}\right)$, and $t_{4}\left(\mathbf{r}\right)$
are relatively shifted by $\frac{l_{1}}{2}\mathbf{e}_{x}$, $\frac{l_{2}}{2}\mathbf{e}_{y}$,
and $\frac{l_{1}}{2}\mathbf{e}_{x}+\frac{l_{2}}{2}\mathbf{e}_{y},$
respectively. Calculating and summing the corresponding Fourier transforms
(using eq. {[}\ref{eq:Tint(w)_app}]) we find for the combined map
that 

\begin{eqnarray}
T_{s}\left(\mathbf{w}\right) & = & T_{1}\left(\mathbf{w}\right)+T_{2}\left(\mathbf{w}\right)+T_{3}\left(\mathbf{w}\right)+T_{4}\left(\mathbf{w}\right)\nonumber \\
 & = & \frac{1}{\left(l_{1}l_{2}\right)^{2}}\sum_{s,t=-\infty}^{\infty}\left[c_{1}+\left(-1\right)^{s}c_{2}+\left(-1\right)^{t}c_{3}+\left(-1\right)^{s+t}c_{4}\right]W\left(\mathbf{w}-\mathbf{w}_{st}\right)T\left(\mathbf{w}-\mathbf{w}_{st}\right),\label{eq:Ts(w)_a}\end{eqnarray}

\noindent where $c_{i}$ is the normalization constant associated
with $t_{i}\left(\mathbf{r}\right)$. Correspondingly, we further
define $\Delta c_{j}=c_{j}-c_{1}$, for $j=2,\,3,\,4$, and we rewrite
equation (\ref{eq:Ts(w)_a}) as

\begin{eqnarray}
T_{s}\left(\mathbf{w}\right) & = & \frac{1}{\left(l_{1}l_{2}\right)^{2}}\sum_{s,t=-\infty}^{\infty}\left\{ 4c_{1}W\left(\mathbf{w}-{2\mathbf{w}}_{st}\right)T\left(\mathbf{w}-2\mathbf{w}_{st}\right)\right.\nonumber \\
 &  & \left.+\left[\left(-1\right)^{s}\Delta c_{2}+\left(-1\right)^{t}\Delta c_{3}+\left(-1\right)^{s+t}\Delta c_{4}\right]W\left(\mathbf{w}-\mathbf{w}_{st}\right)T\left(\mathbf{w}-\mathbf{w}_{st}\right)\right\} .\label{eq:Ts(w)_b}\end{eqnarray}

As will be soon discussed in \S \ref{sub:weight}, the magnitude
of the coefficient $\Delta c_{j}$  depends on the width of the weighting
function $w\left(\mathbf{r}\right)$; the wider the function, the
smaller the coefficient, and vice-versa. As we will see below, there
are good reasons to limit the width of the weighting function, but
if we assume for the moment that $w\left(\mathbf{r}\right)$ is such
that $\Delta c_{j}\simeq0$, then 

\begin{equation}
T_{s}\left(\mathbf{w}\right)\simeq\frac{4c_{1}}{\left(l_{1}l_{2}\right)^{2}}\sum_{s,t=-\infty}^{\infty}W\left(\mathbf{w}-{2\mathbf{w}}_{st}\right)T\left(\mathbf{w}-2\mathbf{w}_{st}\right),\label{eq:Ts(w)_app}\end{equation}

\noindent and

\begin{equation}
t_{s}\left(\mathbf{r}\right)\simeq\left[t\left(\mathbf{r}\right)\otimes w\left(\mathbf{r}\right)\right]\cdot\frac{c_{1}}{l_{1}l_{2}}\sum_{i,k=-\infty}^{\infty}\delta\left(\mathbf{r}-\frac{\mathbf{r}_{ik}}{2}\right).\label{eq:ts(r)_app}\end{equation}

It is instructive to compare this result with the corresponding equation
for an ESG $t^{\prime\prime}\left(\mathbf{r}\right)$, similar to
that of equation (\ref{eq:t'(r)_ESG}) but with half the sampling
interval in each direction 

\begin{eqnarray}
t^{\prime\prime}\left(\mathbf{r}\right) & = & t\left(\mathbf{r}\right)\cdot\sum_{i,k=-\infty}^{\infty}\delta\left(\mathbf{r}-\frac{\mathbf{r}_{ik}}{2}\right)\label{eq:t''(r)_ESG}\\
T^{\prime\prime}\left(\mathbf{w}\right) & = & \frac{4}{l_{1}l_{2}}\sum_{s,t=-\infty}^{\infty}T\left(\mathbf{w}-2\mathbf{w}_{st}\right).\label{eq:T''(w)_ESG}\end{eqnarray}

Again we see that, apart from a multiplication factor, the (approximate)
interpolated grid $t_{s}\left(\mathbf{r}\right)$ differs from $t^{\prime\prime}\left(\mathbf{r}\right)$
by the presence of the convolution by $w\left(\mathbf{r}\right)$.
Although equations (\ref{eq:Ts(w)_app}) and (\ref{eq:ts(r)_app})
are approximations and the aforementioned correspondence between $t_{s}\left(\mathbf{r}\right)$
and $t^{\prime\prime}\left(\mathbf{r}\right)$ may fail for a given
weighting function (i.e., $\Delta c_{j}\neq0$ in general in eq. {[}\ref{eq:Ts(w)_b}]),
this simplification is often reasonable (see \S \ref{sub:weight}).
At any rate, one can more closely approach the idealization of equation
(\ref{eq:ts(r)_app}) by broadening the width of the weighting function
$w\left(\mathbf{r}\right)$ (with a corresponding loss in spatial
resolution, however).

\subsection{Selection of the Weighting Function}

\label{sub:weight}There is a fair amount of subjectiveness in choosing
the specific form and characteristics of a weighting function. We
choose the following two criteria as guidelines for achieving this:

\begin{enumerate}
\item The function must be sufficiently broad so that its amplitude is large
enough at the interpolated positions, while not being too broad to
significantly degrade the resolution of the map.
\item Its spectral extent must be such that it filters out spatial frequencies
for which $\left|u\right|<\left(2l_{1}\right)^{-1}$ or $\left|v\right|<\left(2l_{2}\right)^{-1}$
(see eq. {[}\ref{eq:Tint(w)}] and the discussion that follows).
\end{enumerate}
We now show how this can be practically implemented by considering
the case of SHARC-II (or SHARP) where $\Delta l_{1}=l_{1}=\Delta l_{2}=l_{2}$.
Furthermore, we approximate the SHARC-II PSF with the following Gaussian
profile

\begin{eqnarray*}
h\left(\mathbf{r}\right) & = & \frac{1}{2\pi\sigma^{2}}e^{-\frac{1}{2}\left(\frac{\left|\mathbf{r}\right|}{\sigma}\right)^{2}}\\
H\left(\mathbf{w}\right) & = & e^{-2\pi^{2}\sigma^{2}\left|\mathbf{w}\right|^{2}}\equiv e^{-\frac{1}{2}\left(\frac{\left|\mathbf{w}\right|}{\Sigma}\right)^{2}}.\end{eqnarray*}

The PSF size is usually defined by its full-width-half-magnitude (FWHM),
which is approximately 9 arcseconds for SHARC-II at 350 $\mu$m \citep{Dowell 2003}.
This gives $\sigma=\mathrm{FWMH/\sqrt{8\ln\left(2\right)}}\simeq3.8$
arcseconds; we will use standard deviations to specify widths of Gaussian
PSFs. With this definition, the one-sided bandwidth associated with
SHARC-II at 350 $\mu$m is

\[
\Sigma=\frac{1}{2\pi\sigma}\simeq\frac{1}{23.9}\,\,\mathrm{arcseconds}^{-1}.\]

Since $l_{1}=l_{2}\simeq4.7$ arcseconds, then $\Sigma^{-1}>2l_{1}$
and the Nyquist sampling criterion is met, as previously assumed.
If we were to choose the weighting function $w\left(\mathbf{r}\right)$
to also be Gaussian, and of width $\varpi$, then to satisfy Criterion
2 above we must have

\[
\varpi\gtrsim\frac{l_{1}}{\pi}\simeq1.5\,\,\mathrm{arcseconds.}\]

Taking the lower limit for the size of the kernel, we can evaluate
the new resolution of the map with

\begin{equation}
\sqrt{\sigma^{2}+\varpi^{2}}=\sigma\sqrt{1+\frac{l_{1}^{2}}{\pi^{2}\sigma^{2}}}\simeq1.07\sigma=4.1\,\,\mathrm{arcseconds,}\label{eq:beamwidth}\end{equation}

\noindent which corresponds to an equivalent PSF width of 9.6 arcseconds
for SHARC-II and a loss of approximately 7\% in spatial resolution.
Finally, the relative amplitude of the weighting function at a distance
of one-half pixel away from the position of interpolation would be

\begin{equation}
2\pi\varpi^{2}w\left(\mathbf{r}\right)\left|_{r=\frac{l_{1}}{2}}\right.=e^{-\frac{\pi^{2}}{8}}=0.29.\label{eq:weigth}\end{equation}

\begin{figure}
\epsscale{0.7}
\plotone{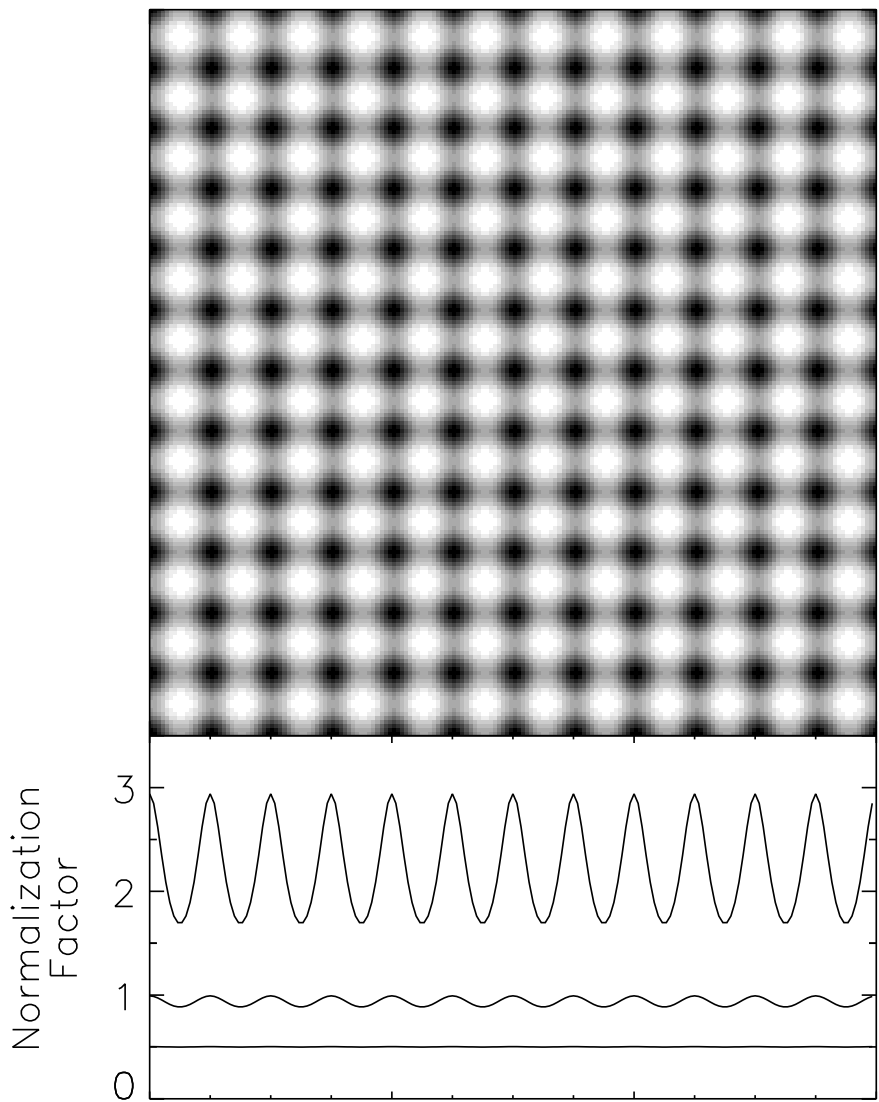}

\caption{\label{fig:normalization}A map (top) of the normalization function
$n\left(\mathbf{r}\right)$ for a SHARP ESG with a weighting function
with $\varpi=l_{1}/\pi$ ($l_{1}=l_{2}\simeq4.7$ arcseconds for SHARP).
The top most curve in the lower part of the figure is a cut through
a row or column of pixels for the normalization map. The bottom two
curves are similar cuts for weighting functions of $\varpi=1.3\, l_{1}/\pi$
and $l_{1}/\sqrt{\pi}\simeq1.8\, l_{1}/\pi$, respectively.}
\end{figure}

\begin{figure}
\plotone{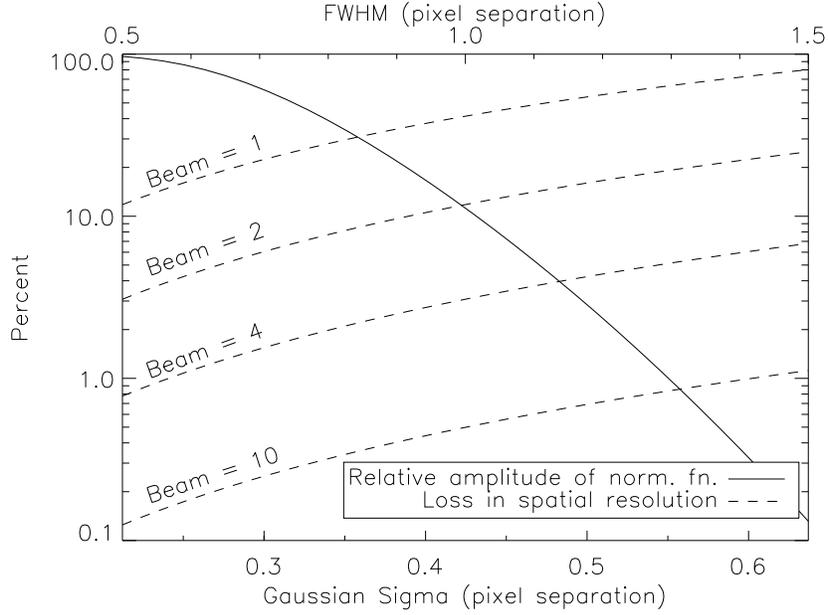}

\caption{\label{fig:ripple}Trends in normalization function (solid line) and
spatial resolution degradation (dashed lines) with smoothing kernel
size. The kernel sizes on the abscissa are given as the Gaussian width
($\varpi$; lower axis) and FWHM ($=\sqrt{8\ln\left(2\right)}\,\varpi$;
upper axis), both in units of the array pixel separation. The amplitude
of the normalization function's spatial variation is given as a percentage
of the function's average value (see Fig. \ref{fig:normalization}).
The corresponding loss in spatial resolution is described by equation
(\ref{eq:beamwidth}) and the following text. This is plotted here
for different beam sizes and indicated in units of pixel separation.
For example, the case of SHARP, with $\mathrm{Beam}=9\arcsec/4.7\arcsec\approx2$,
closely corresponds to the second dashed curve (from the top). }
\end{figure}

Although equations (\ref{eq:beamwidth}) and (\ref{eq:weigth}) satisfy
Criterion 1 above and one could reasonably choose the corresponding
weighting function to interpolate a map, one should nonetheless verify
that the coefficients $\Delta c_{j}$ resulting from the interpolation
process (see \S \ref{sub:Interpolation}) are sufficiently small
when seeking to increase the density of samples in the final grid.
Doing so will ensure that the approximation leading to equation (\ref{eq:ts(r)_app})
is adequate, for example. Figure \ref{fig:normalization} shows a
map (top) of the normalization function $n\left(\mathbf{r}\right)$
for a SHARP ESG with the lower-limit weighting function considered
above ($\varpi=l_{1}/\pi\simeq1.5$ arcseconds). The top most curve
in the lower part of the figure is a cut through a row or column of
pixels for the normalization map. The bottom two curves are similar
cuts for weighting functions of $\varpi=1.3\, l_{1}/\pi$ and $l_{1}/\sqrt{\pi}\simeq1.8\, l_{1}/\pi$,
respectively. It is clear from these curves that the relative amplitude
of the $\Delta c_{j}$ coefficients, which can be asserted from the
level of ripple on the curves, exhibits a strong dependency on the
width of the weighting function. For example, the two larger weighting
functions in the lower part of Figure \ref{fig:normalization} exhibit
variations in amplitude of 11\% and 1\% for a loss in spatial resolution
of 13\% and 22\%, respectively. This behavior is traced to the fact
that a larger weighting function will more completely cover the space
located between neighboring sampling positions, hence the existence
of a smoother normalization function $n\left(\mathbf{r}\right)$.
This will be better visualized for the general case with the graph
shown in Figure \ref{fig:ripple} where trends in normalization function
(solid line) and spatial resolution degradation (dashed lines) with
smoothing kernel size are plotted (see the corresponding caption).

\section{The Regularly Sampled Grid}

\label{sec:RSG}
\begin{figure}
\plotone{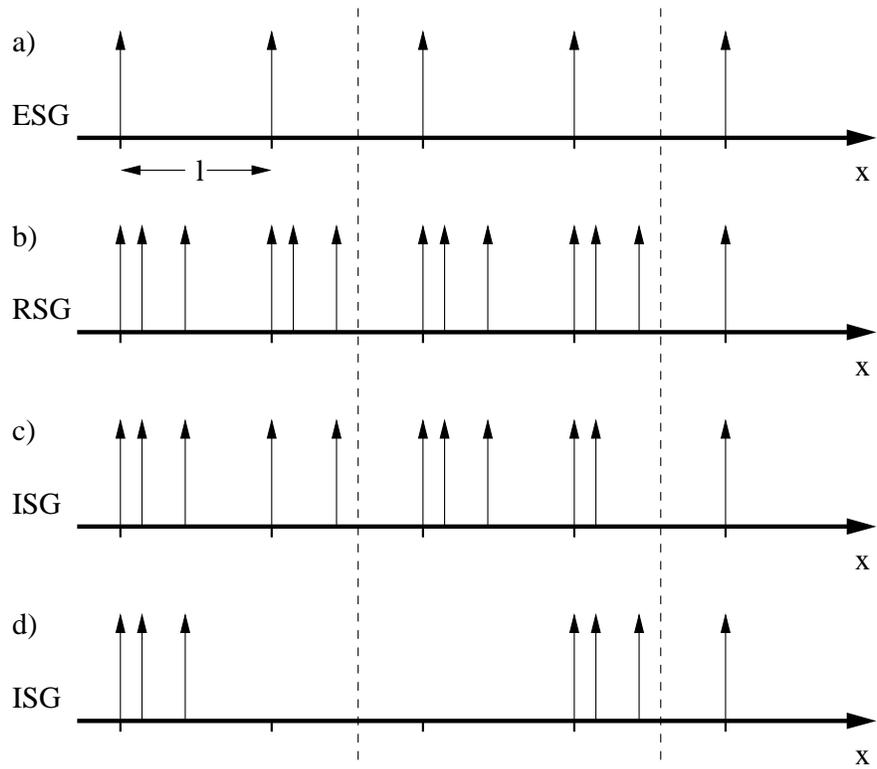}

\caption{\label{fig:grids}One-dimensional examples of (a) an Evenly Sampled
Grid (ESG), (b) a Regularly Sampled Grid (RSG), and (c) and (d) two
Irregularly Sampled Grids (ISGs) are shown. Interpolations at the
positions of the two vertical broken lines would require weighting
functions that have a common normalization factor for the ESG and
RSG, but different normalization factors for the ISGs. Dirac distributions
are shown as vertical arrows.}
\end{figure}

We define a Regularly Sampled Grid (RSG) as being
a generalization of the ESG discussed in the previous section. That
is, a RSG has a well defined periodicity (just as the ESG), but it
is a grid for which the pattern of Dirac distributions is more complex.
While along a coordinate axis of the ESG there is only one Dirac distribution
for a given period, a RSG may have many Dirac distributions (not necessarily
evenly spaced) over the same interval. This difference is illustrated
in cases (a) and (b) of Figure \ref{fig:grids} for one-dimensional
versions of an ESG and a RSG, respectively. Practically speaking,
a RSG would be encountered any time that maps of similar characteristics,
but translated relative to one another, are combined together to form
a unique and final map. An example for this would be astronomical
images of a given object at different pointing positions. 

It should be apparent from this discussion and Figure \ref{fig:grids}\emph{b}
that a RSG, which we again denote by $t^{\prime}\left(\mathbf{r}\right)$,
can be simply expressed as a combination of a set of relatively displaced
ESGs with

\begin{equation}
t^{\prime}\left(\mathbf{r}\right)=t\left(\mathbf{r}\right)\cdot\sum_{p=1}^{n_{g}}\left[\sum_{i,k=-\infty}^{\infty}\delta\left(\mathbf{r}-\mathbf{r}_{ik}-\mathbf{d}_{p}\right)\right],\label{eq:t'_RSG}\end{equation}

\noindent where $t\left(\mathbf{r}\right)$ is defined in equation
(\ref{eq:t(r)_conv}) and

\begin{equation}
\mathbf{d}_{p}\left(\mathbf{r}\right)=x_{p}\mathbf{e}_{x}+y_{p}\mathbf{e}_{y}\label{eq:d_p}\end{equation}

\noindent is the relative displacement associated with the $p$th
of the $n_{g}$ ESGs that make up the RSG. It is straightforward to
calculate the Fourier transform of equation (\ref{eq:t'_RSG}) to
get

\begin{equation}
T^{\prime}\left(\mathbf{w}\right)=\frac{1}{l_{1}l_{2}}\sum_{p=1}^{n_{g}}\sum_{m,n=-\infty}^{\infty}T\left(\mathbf{w}-\mathbf{w}_{mn}\right)e^{-j2\pi\mathbf{w}_{mn}\cdot\mathbf{d}_{p}},\label{eq:T'(w)_RSG}\end{equation}

\noindent with $\mathbf{w}_{mn}$ defined in equation (\ref{eq:w_mn}).

\noindent The important aspect to emphasize for the interpolation
of a RSG is that, as was the case for an ESG, the normalization factor
$n\left(\mathbf{r}\right)$ in equation (\ref{eq:z_int(r)}) is common
to all interpolated points similarly located within a one-period segment
anywhere on the grid. This is illustrated with the vertical broken
lines in Figure \ref{fig:grids}. The existence of such a common normalization
factor could be effectively adopted as the definition for a RSG.

If we interpolate our RSG using a weighting function $w\left(\mathbf{r}\right)$
that satisfies Criterion 2 above, then the high spatial frequency
components of the spectrum will be filtered out. Therefore, starting
from equation (\ref{eq:T'(w)_RSG}), and using steps similar to those
that led from equation (\ref{eq:T'(w)}) to equations (\ref{eq:Tint(w)})
and (\ref{eq:Tint(w)_app}), the resulting interpolated grid becomes

\begin{equation}
T_{\mathrm{int}}\left(\mathbf{w}\right)=\frac{n_{g}c}{\left(l_{1}l_{2}\right)^{2}}\sum_{s,t=-\infty}^{\infty}W\left(\mathbf{w}-\mathbf{w}_{st}\right)T\left(\mathbf{w}-\mathbf{w}_{st}\right)e^{-j2\pi\mathbf{w}_{st}\cdot\mathbf{a}_{pq}},\label{eq:Tint(w)_RSG}\end{equation}

\noindent and

\begin{equation}
t_{\mathrm{int}}\left(\mathbf{r}\right)=\left[t\left(\mathbf{r}\right)\otimes w\left(\mathbf{r}\right)\right]\cdot\frac{n_{g}c}{l_{1}l_{2}}\sum_{i,k=-\infty}^{\infty}\delta\left(\mathbf{r}-\mathbf{r}_{ik}-\mathbf{a}_{pq}\right).\label{eq:tint(r)_RSG}\end{equation}

Once again $c$ and $\mathbf{a}_{pq}$ denote, respectively, the common
normalization factor and the displacement of the new interpolated
grid in relation to the original grid (see eq. {[}\ref{eq:a_pq}]).
Obviously, the same comments apply for the map resulting from the
re-sampling of a RSG here as for an ESG in \S \ref{sec:ESG}. That
is, the most notable effect of the interpolation/re-sampling process
is the presence of the convolution by the weighting function in equation
(\ref{eq:tint(r)_RSG}).

\section{The Irregularly Sampled Grid}

\label{sec:ISG}An Irregularly Sampled Grid (ISG) can manifest itself
in different ways. For example, Figure \ref{fig:grids}\emph{c} shows
a case where the distribution of Dirac functions within a given base
period (of length $l$ in the figure) is not the same from one interval
to the next. Another possibility is shown in Figure \ref{fig:grids}\emph{d}
where no Dirac distributions are present for some intervals. This
can be likened to situations where pixels are missing from an array
detector (see below).

The problem in the analysis of an ISG is twofold. First, there is
no simple way of expressing the Fourier transform of irregularly spaced
Dirac distributions such that the spectrum will show a repeating pattern
of some frequency, as is the case for an ESG or a RSG. Moreover, the
lack of regularity in the positions of the Dirac distributions implies
that there does not exist a common normalization factor when performing
interpolations to create an ESG from an ISG (see below). Nevertheless,
it is still possible to analyze some specific types of ISGs. We deal
with two possible cases in what follows.

\subsection{The Combination of Relatively Translated and Rotated ESGs}

\begin{figure}
\epsscale{0.7}\plotone{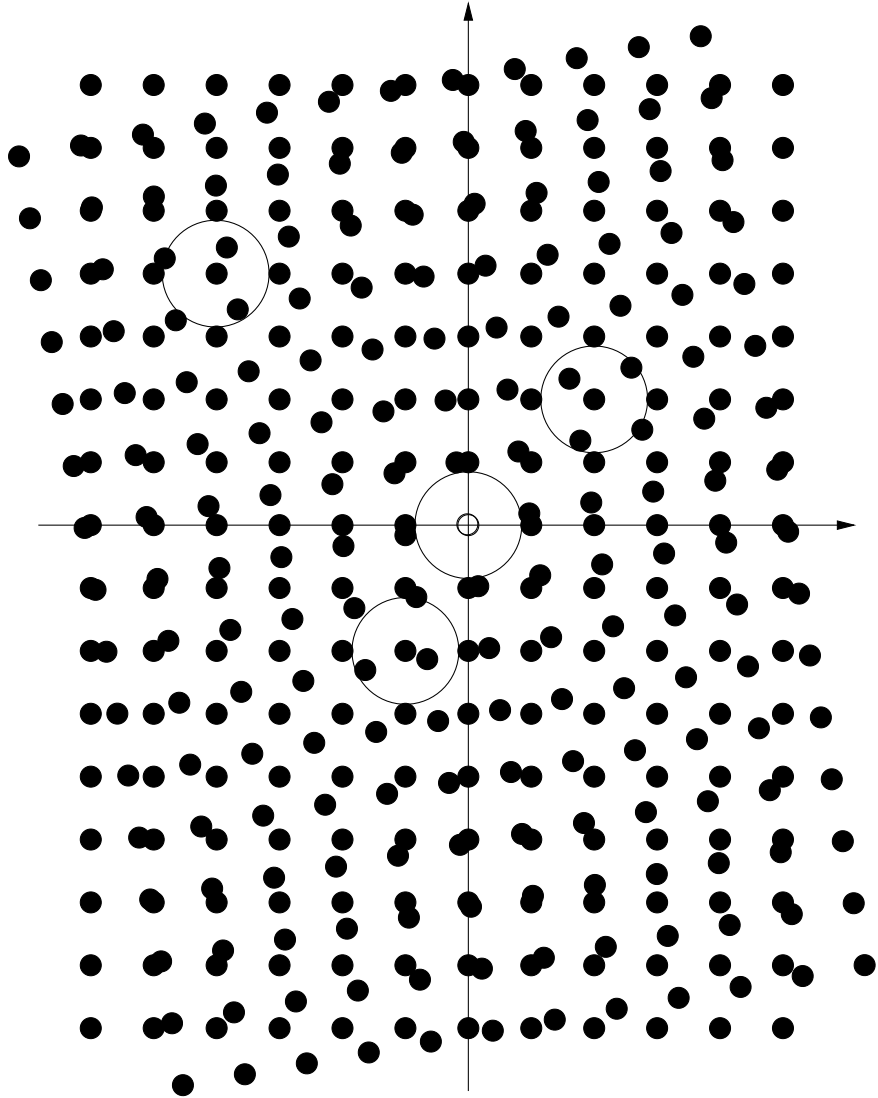}

\caption{\label{fig:ISG_rotation}A combination of two relatively rotated ESGs.
Every small dark dot corresponds to a Dirac distribution, and the
position of the small empty circle is the origin of the maps and of
the rotation for one of the two grids. Its relative rotation is of
10 degrees with respect to the coordinate axes (also shown). Neither
ESG is translated. The four large empty circles correspond to the
footprint of a predetermined weighting function.}
\end{figure}

It often happens that an astronomical source will be observed at different
times, when it is at different locations and orientations on the celestial
sphere. Invariably, we seek to combine the resulting images to form
a final map of the object. If the array detector (which we assume
perfect and therefore able to generate ESGs of data) used to record
the images is part of an instrument that is unable to precisely track
the apparent rotation of the source on the sky, then the different
images of the source will be sampled with ESGs that will be rotated
and possibly translated relative to each other. A simple example is
shown in Figure \ref{fig:ISG_rotation} where two ESGs are combined:
one rotated by 10 degrees with respect to the other. These grids are
not relatively translated (see the caption).

Perhaps the most important aspect of Figure \ref{fig:ISG_rotation}
is the fact that the combination of the two ESGs produces a grid which
has an irregular pattern of Dirac distributions, as can be asserted
by the coverage of a predetermined weighting function (shown with
large empty circles in the figure). Clearly, any weighting function
$w\left(\mathbf{r}\right)$ is likely to cover a different number
of samples at different positions on the map. The main consequence
resulting from this fact will be the absence of a common normalizing
factor at the different locations where interpolations are performed
(note that the normalization function in eq. {[}\ref{eq:n(r)_ESG}]
will not be periodic). We can therefore expect that interpolated maps
originating from ISGs will be more complex than those resulting from
ESGs and RSGs. 

Another point to consider is the possible relative rotation between
the different maps to be combined. Because of this, we will do well
to use the fact that the Fourier transform of a rotated map is the
rotated Fourier transform of the original (i.e., without rotation)
map. That is, if we have the Fourier pair

\[
g\left(\mathbf{r}\right)\Leftrightarrow G\left(\mathbf{w}\right),\]

\noindent then it is also true that (see Appendix)

\[
g\left(\mathbf{Rr}\right)\Leftrightarrow G\left(\mathbf{Rw}\right),\]

\noindent where $\mathbf{R}$ stands for the rotation operation (i.e.,
matrix). Because of this property of the Fourier transform we can
express an ISG $t^{\prime}\left(\mathbf{r}\right)$ composed of $n_{g}$
rotated and translated ESGs and its Fourier transform as

\begin{eqnarray}
t^{\prime}\left(\mathbf{r}\right) & = & t\left(\mathbf{r}\right)\cdot\sum_{p=1}^{n_{g}}\sum_{i,k=-\infty}^{\infty}\delta\left(\mathbf{r}-\mathbf{R}_{p}\mathbf{r}_{ik}-\mathbf{d}_{p}\right)\label{eq:t'(r)_ISG}\\
T^{\prime}\left(\mathbf{w}\right) & = & \frac{1}{l_{1}l_{2}}\sum_{p=1}^{n_{g}}\sum_{m,n=-\infty}^{\infty}T\left(\mathbf{w}-\mathbf{R}_{p}\mathbf{w}_{mn}\right)e^{-j2\pi\mathbf{w}_{mn}\cdot\mathbf{d}_{p}},\label{eq:T'(w)_ISG}\end{eqnarray}

\noindent where $\mathbf{R}_{p}$ and $\mathbf{d}_{p}=x_{p}\mathbf{e}_{x}+y_{p}\mathbf{e}_{y}$
are, respectively, the rotation matrix and the translation vector
corresponding to grid $p$. Just as for the RSG our goal is to generate
an ESG $t_{\mathrm{int}}\left(\mathbf{r}\right)$ of periods $l_{1}$
and $l_{2}$ (along the $x$ and $y$ axes, respectively) from $t^{\prime}\left(\mathbf{r}\right)$.
Although, as was previously pointed out, we cannot express the interpolation
process with a simple convolution with a weighting function $w\left(\mathbf{r}\right)$,
we can still use the general expression given in equation (\ref{eq:tint(r)}).
That is, with $\mathbf{a}_{pq}$ denoting the origin of the new interpolated
grid (see eq. {[}\ref{eq:a_pq}]) we have

\begin{eqnarray}
t_{\mathrm{int}}\left(\mathbf{r}\right) & = & \left[t^{\prime}\left(\mathbf{r}\right)\otimes w\left(\mathbf{r}\right)\right]\cdot n\left(\mathbf{r}\right)\sum_{i,k=-\infty}^{\infty}\delta\left(\mathbf{r}-\mathbf{r}_{ik}-\mathbf{a}_{pq}\right)\nonumber \\
 & = & \left\{ \left[t\left(\mathbf{r}\right)\cdot\sum_{p=1}^{n_{g}}\sum_{i,k=-\infty}^{\infty}\delta\left(\mathbf{r}-\mathbf{R}_{p}\mathbf{r}_{ik}-\mathbf{d}_{p}\right)\right]\otimes w\left(\mathbf{r}\right)\right\} \nonumber \\
 &  & \,\,\,\cdot n\left(\mathbf{r}\right)\sum_{i,k=-\infty}^{\infty}\delta\left(\mathbf{r}-\mathbf{r}_{ik}-\mathbf{a}_{pq}\right).\label{eq:tint(r)_ISG}\end{eqnarray}

Calculating the Fourier transform of equation (\ref{eq:tint(r)_ISG})
we get

\begin{eqnarray*}
T_{\mathrm{int}}\left(\mathbf{w}\right) & = & \left(\left\{ \left[\frac{1}{l_{1}l_{2}}\sum_{p=1}^{n_{g}}\sum_{s,t=-\infty}^{\infty}T\left(\mathbf{w}-\mathbf{R}_{p}\mathbf{w}_{st}\right)e^{-j2\pi\mathbf{w}_{st}\cdot\mathbf{d}_{p}}\right]W\left(\mathbf{w}\right)\right\} \otimes N\left(\mathbf{w}\right)\right)\\
 &  & \,\,\,\otimes\frac{1}{l_{1}l_{2}}\sum_{m,n=-\infty}^{\infty}\delta\left(\mathbf{w}-\mathbf{w}_{mn}\right)e^{-j2\pi\mathbf{w}_{mn}\cdot\mathbf{a}_{pq}},\end{eqnarray*}

\noindent which, using the assumption that $W\left(\mathbf{w}\right)$
is such that it filters out the higher frequency components of the
spectrum, can be approximated to

\begin{equation}
T_{\mathrm{int}}\left(\mathbf{w}\right)=\frac{n_{g}}{\left(l_{1}l_{2}\right)^{2}}\sum_{m,n=-\infty}^{\infty}e^{-j2\pi\mathbf{w}_{mn}\cdot\mathbf{a}_{pq}}\left\{ \left[T\left(\mathbf{w}\right)W\left(\mathbf{w}\right)\right]\otimes N\left(\mathbf{w}\right)\right\} _{\mathbf{w}=\mathbf{w}-\mathbf{w}_{mn}}.\label{eq:Tint(w)_ISG_approx}\end{equation}

Correspondingly, we can approximate equation (\ref{eq:tint(r)_ISG})
to

\begin{equation}
t_{\mathrm{int}}\left(\mathbf{r}\right)=\left\{ \left[t\left(\mathbf{r}\right)\otimes w\left(\mathbf{r}\right)\right]\cdot n\left(\mathbf{r}\right)\right\} \cdot\frac{n_{g}}{l_{1}l_{2}}\sum_{i,k=-\infty}^{\infty}\delta\left(\mathbf{r}-\mathbf{r}_{ik}-\mathbf{a}_{pq}\right).\label{eq:tint(r)_ISG_approx}\end{equation}

Equation (\ref{eq:Tint(w)_ISG_approx}) shows best the effect of the
lack of a common normalization factor on the interpolation process.
Since the Fourier transform $N\left(\mathbf{w}\right)$ of the normalization
function is convolved with the weighted (and low-pass filtered) spectrum
$T\left(\mathbf{w}\right)W\left(\mathbf{w}\right)$, the resulting
spectrum of the interpolated ESG $t_{\mathrm{int}}\left(\mathbf{r}\right)$
is broadened by $N\left(\mathbf{w}\right)$. It is interesting to
note that $w\left(\mathbf{r}\right)$ and $n\left(\mathbf{r}\right)$
have opposite effects on the signal. That is, the weighting function
restricts the extent of the spectrum, while the normalization function
extends it. 

\begin{figure}
\plotone{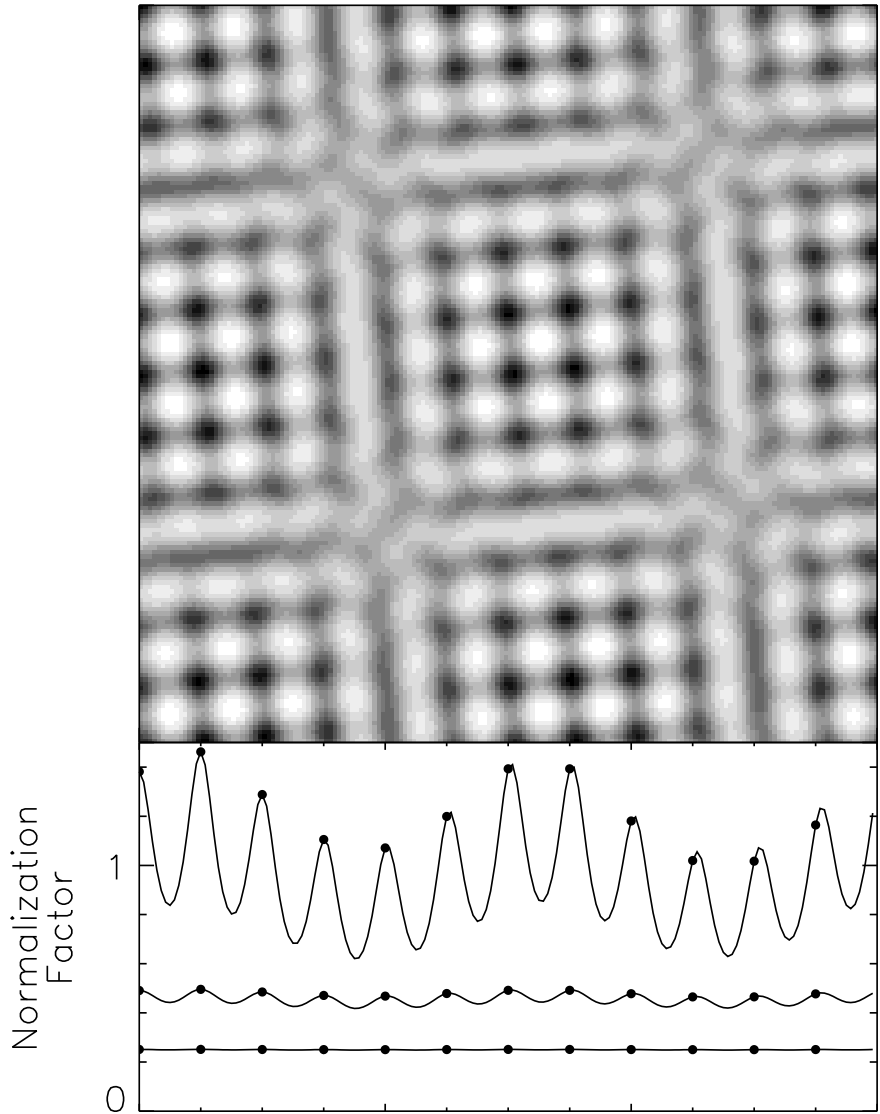}

\caption{\label{fig:ISG_ripple}A combination of two relatively rotated ESGs,
as those of Figure \ref{fig:ISG_rotation}. We show a map (top) of
the normalization function $n\left(\mathbf{r}\right)$ for the resulting
SHARP ISG using a weighting function with $\varpi=l_{1}/\pi$ ($l_{1}=l_{2}\simeq4.7$
arcseconds for SHARP). The top most curve in the lower part of the
figure is an arbitrary cut through a row of pixels for this normalization
map. The bottom two curves are similar cuts for weighting functions
of $\varpi=1.3\, l_{1}/\pi$ and $l_{1}/\sqrt{\pi}\simeq1.8\, l_{1}/\pi$,
respectively. The black dots highlights the values taken by $n\left(\mathbf{r}\right)$
for a re-sampling onto an ESG at the original sampling rate. It is
clear from the top two cuts that the normalization factor is not constant
in general.}
\end{figure}

Given such an ISG, it should be in principle possible to evaluate
$n\left(\mathbf{r}\right)$ and quantify its effect. In particular,
one should ensure that the two previous criteria (see \S \ref{sub:weight})
used to select the weighting function are met. Optimally, $n\left(\mathbf{r}\right)$
will be sufficiently slowly varying, and of low enough amplitude,
that the spectral broadening will be minimal. To make this clearer,
we show in Figure \ref{fig:ISG_ripple} an example consisting of a
combination of two relatively rotated ESGs, similar to those of Figure
\ref{fig:ISG_rotation} (i.e., one rotated by 10 degrees with respect
to the other, and no relative translation between the two). The map
at the top of the figure is for the normalization function $n\left(\mathbf{r}\right)$
of the resulting SHARP ISG using a weighting function with $\varpi=l_{1}/\pi$
($l_{1}=l_{2}\simeq4.7$ arcseconds for SHARP). The top most curve
in the lower part of the figure is an arbitrary cut through a row
of pixels for this normalization map. The bottom two curves are similar
cuts for weighting functions of $\varpi=1.3\, l_{1}/\pi$ and $l_{1}/\sqrt{\pi}\simeq1.8\, l_{1}/\pi$,
respectively. The black dots highlight the values taken by $n\left(\mathbf{r}\right)$
for a re-sampling onto an ESG at the original sampling rate. It is
clear from the top two cuts that the normalization factor is not constant
in general. One can also assert from this that the spectrum due to
a broad source (in relation to the size of the map) would be significantly
more broadened by the weighting function that produced the top curve
(i.e., with $\varpi=l_{1}/\pi$) than by the other two.

\subsection{The Effects of Missing Samples}

\label{sub:missing}It is a common, if unfortunate, fact that detectors
arrays used in astronomy will often contain pixels that are either
performing significantly below specifications or are completely unusable.
Astronomers usually work around the difficulties occasioned by these
so-called {}``missing'' pixels by dithering the array during observations,
thus ensuring complete mapping of the source under study. It would
be, however, instructive to analyze and quantify the impact that missing
pixels would have on the representation of astronomical signals without
such corrective techniques. 

Although we will take into account the fact that the map obtained
from the array is composed of a finite number of samples, our approach
will consist of first temporarily lending it an infinite character
and then removing it. More precisely, although the size of the (rectangular)
detector array considered in this section is $N_{1}l_{1}\times N_{2}l_{2}$,
we first assume that the two-dimensional pattern of $N_{1}N_{2}$
pixels (including the missing pixels) repeats infinitely in all directions.
The underlying assumption is that the finiteness of the map will be
restored in the end by windowing with the appropriate aperture function.
Using this approach we express the sampled signal (before windowing)
as

\begin{equation}
t^{\prime}\left(\mathbf{r}\right)=t\left(\mathbf{r}\right)\cdot\sum_{\mathrm{pix}}\sum_{s,t=-\infty}^{\infty}\delta\left(\mathbf{r}-\mathbf{d}_{st}-\mathbf{r}_{\mathrm{pix}}\right),\label{eq:t'(r)_miss}\end{equation}

\noindent with $\mathbf{r}_{\mathrm{pix}}$ the position of a pixel
on the array and (take note of the periods)

\begin{equation}
\mathbf{d}_{st}=sN_{1}l_{1}\mathbf{e}_{x}+tN_{2}l_{2}\mathbf{e}_{y}.\label{eq:r_mn_miss}\end{equation}

That is, we first account for the pixels of the finite array through
the summation $\sum_{\mathrm{pix}}$, and then associate a Dirac train
of periods $\left(N_{1}l_{1},N_{2}l_{2}\right)$ to each pixel. These
Dirac trains are relatively translated in space and are accounted
for by the summations on $s$ and $t$ in equation (\ref{eq:t'(r)_miss}).
The Fourier transform of the combined Dirac trains is

\begin{equation}
\sum_{\mathrm{pix}}\sum_{s,t=-\infty}^{\infty}\delta\left(\mathbf{r}-\mathbf{d}_{st}-\mathbf{r}_{\mathrm{pix}}\right)\Leftrightarrow\left[\sum_{\mathrm{pix}}e^{-j2\pi\mathbf{w}\cdot\mathbf{r}_{\mathrm{pix}}}\right]\cdot\frac{1}{N_{1}N_{2}l_{1}l_{2}}\sum_{m,n=-\infty}^{\infty}\delta\left(\mathbf{w}-\mathbf{w}_{mn}\right),\label{eq:Dtrain_miss}\end{equation}

\noindent with

\begin{equation}
\mathbf{w}_{mn}=\frac{m}{N_{1}l_{1}}\mathbf{e}_{x}+\frac{n}{N_{2}l_{2}}\mathbf{e}_{y}.\label{eq:w_mn_miss}\end{equation}

Note that the minimum separation between two Dirac distributions in
frequency space is $1/Nl$, where $Nl$ is the greater of $N_{1}l_{1}$
and $N_{2}l_{2}$. We write the right hand side of equation (\ref{eq:Dtrain_miss}),
which we denote by $D\left(\mathbf{w}\right)$, as follows

\begin{equation}
D\left(\mathbf{w}\right)=\frac{1}{l_{1}l_{2}}\sum_{m,n=-\infty}^{\infty}E\left(\mathbf{w}_{mn}\right)\delta\left(\mathbf{w}-\mathbf{w}_{mn}\right),\label{eq:D(w)_miss}\end{equation}

\noindent with

\begin{equation}
E\left(\mathbf{w}_{mn}\right)=\frac{1}{N_{1}N_{2}}\sum_{\mathrm{pix}}e^{-j2\pi\mathbf{w}_{mn}\cdot\mathbf{r}_{\mathrm{pix}}}.\label{eq:E(w)_miss}\end{equation}

The effect of missing pixels can now easily be quantified, at least
in principle, by omitting them from the $\sum_{\mathrm{pix}}$ summation
in equations (\ref{eq:E(w)_miss}). As an example, consider the case
where the $M\times M$ pixels in the {}``top-right'' corner of the
array are missing. For this particular example, we will do well to
make the following substitutions 

\begin{eqnarray*}
\sum_{\mathrm{pix}} & \rightarrow & \sum_{i}\sum_{k}\\
\mathbf{r}_{\mathrm{pix}} & \rightarrow & il_{1}\mathbf{e}_{x}+kl_{2}\mathbf{e}_{y},\end{eqnarray*}

\noindent where $i$ and $k$ indices stand for the columns and rows
of the detector, respectively. We can then transform equation (\ref{eq:E(w)_miss})
to

\begin{eqnarray}
E\left(\mathbf{w}_{mn}\right) & = & \frac{1}{N_{1}N_{2}}\left(\sum_{i=0}^{N_{1}-1}e^{-j2\pi im/N_{1}}\cdot\sum_{k=0}^{N_{2}-M-1}e^{-j2\pi kn/N_{2}}\right.\nonumber \\
 &  & \left.+\sum_{i=0}^{N_{1}-M-1}e^{-j2\pi im/N_{1}}\cdot\sum_{k=N_{2}-M}^{N_{2}-1}e^{-j2\pi kn/N_{2}}\right)\nonumber \\
 & = & \frac{e^{j\pi n\left(M+1\right)/N_{2}}}{N_{1}N_{2}}\left\{ \left(-1\right)^{m+n}e^{j\pi m/N_{1}}\frac{\sin\left(\pi m\right)}{\sin\left(\pi m/N_{1}\right)}\frac{\sin\left[\pi n\left(N_{2}-M\right)/N_{2}\right]}{\sin\left(\pi n/N_{2}\right)}\right.\nonumber \\
 &  & \left.+\left(-1\right)^{m}e^{j\pi m\left(M+1\right)/N_{1}}\frac{\sin\left[\pi m\left(N_{1}-M\right)/N_{1}\right]}{\sin\left(\pi m/N_{1}\right)}\frac{\sin\left(\pi nM/N_{2}\right)}{\sin\left(\pi n/N_{2}\right)}\right\} .\label{eq:E(w)_M^2}\end{eqnarray}

When $M=0$ (i.e., when all the pixels are accounted for) equation
(\ref{eq:E(w)_M^2}) simplifies to

\begin{eqnarray*}
E\left(\mathbf{w}_{mn}\right) & = & \left\{ \begin{array}{cc}
1, & m=m^{\prime}N_{1}\,\mathrm{and}\, n=n^{\prime}N_{2}\\
0, & \mathrm{elsewhere}\end{array}\right.\end{eqnarray*}

\noindent with $m^{\prime}$ and $n^{\prime}$ some integer numbers.
That is to say, equation (\ref{eq:D(w)_miss}) then becomes

\[
D\left(\mathbf{w}\right)=\frac{1}{l_{1}l_{2}}\sum_{m,n=-\infty}^{\infty}\delta\left[\mathbf{w}-\mathbf{w}_{\left(N_{1}m\right)\left(N_{2}n\right)}\right],\]

\noindent which is, as it should be, the same result as was obtained
earlier with equation (\ref{eq:dtrain_ft}) for the ESG.

Although it is necessary to plot $E\left(\mathbf{w}_{mn}\right)$
to assess the effect of an arbitrary distribution of missing pixels,
it should now be clear from equation (\ref{eq:E(w)_M^2}) that its
amplitude is in general non-zero for all values of $m$ and $n$.
In other words, by removing even only one pixel we went from a case
where we only had Dirac functions at frequency intervals of $l_{1}^{-1}$
and $l_{2}^{-1}$ to a situation where they are separated by intervals
of only $\left(N_{1}l_{1}\right)^{-1}$ and $\left(N_{2}l_{2}\right)^{-1}$.
It is important, however, to quantify the relative magnitude of these
Dirac distributions. For example, returning to equation (\ref{eq:E(w)_M^2})
pertaining to our case of the $M\times M$ missing {}``top-right''
pixels, we find that

\begin{eqnarray}
\left|\frac{E\left(\mathbf{w}_{10}\right)}{E\left(\mathbf{w}_{00}\right)}\right| & = & \frac{M\left|\sin\left[\pi\left(N_{1}-M\right)/N_{1}\right]/\sin\left(\pi/N_{1}\right)\right|}{N_{1}N_{2}-M^{2}}\nonumber \\
 & \simeq & \frac{M^{2}}{N_{1}N_{2}-M^{2}},\,\,\,\,\,\,\,\,\mathrm{for}\, M\ll N_{1}.\label{eq:E_10/E_00}\end{eqnarray}

\begin{figure}
\epsscale{0.65}
\rotatebox{270}{\plotone{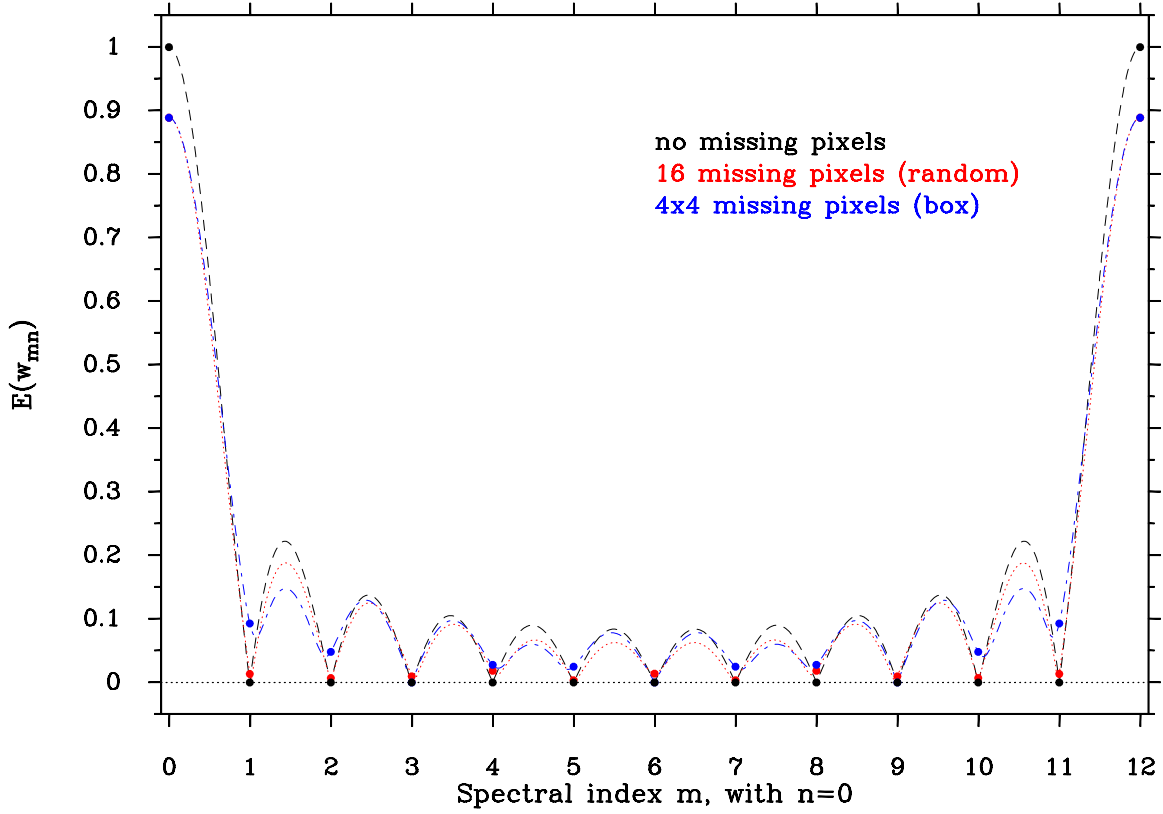}}

\caption{\label{fig:missing}The function $E\left(\mathbf{w}_{mn}\right)$
for $n=0$ when no pixels (black), 16 randomly positioned pixels (red),
and the $4\times4$ {}``top-right'' corner pixels (blue) are missing.
Contrary to the case of an ESG (corresponding to the black dots) where
$E\left(\mathbf{w}_{m0}\right)=0$ when $\left|m\right|\neq0,12,...$,
an ISG (red and blue dots) will in general have $E\left(\mathbf{w}_{mn}\right)\neq0$
for all $m$ and $n$. The amplitude of $E\left(\mathbf{w}_{mn}\right)$
for the relevant (i.e., integer) values of $m$ are shown by the colored
dots. The curves, which include computations at intermediate values
of $m$, are only shown to emphasize the fact that $E\left(\mathbf{w}_{mn}\right)$
can be interpreted as a mask function (see text).}
\end{figure}

This last relation yields the perhaps intuitive result that when only
a small number of pixels are missing the amount of contamination determined
by the ratio expressed in equation (\ref{eq:E_10/E_00}) is approximately
equal to the ratio of the number of missing pixels to that of good
pixels. We should, however, resist the temptation to generalize this
result, since different distributions of $M^{2}$ missing pixels would
give different levels of contamination. Especially if they are not
concentrated in one part of the array, as is the case here. An example
is shown in Figure \ref{fig:missing} where the function $E\left(\mathbf{w}_{mn}\right)$
is plotted for three different cases. Starting with the nominal SHARP
$12\times12$ array, $E\left(\mathbf{w}_{mn}\right)$ is shown for
$n=0$ when \emph{i)} no pixels (black curve and dots), \emph{ii)}
16 randomly positioned pixels (red curve and dots), and \emph{iii)}
the $4\times4$ {}``top-right'' corner pixels (blue curve and dots)
are missing. Contrary to the case of an ESG (corresponding to the
black dots) where $E\left(\mathbf{w}_{mn}\right)=0$ when $\left|m\right|\neq0,12,...$,
an ISG (red and blue dots) will in general have $E\left(\mathbf{w}_{mn}\right)\neq0$
for all $m$ and $n$. It should be clear that $E\left(\mathbf{w}_{mn}\right)$
acts as a mask that will or will not allow the appearance of Dirac
distributions that are more closely spaced in frequency depending
on whether or not there are missing pixels in the array. This serves
to emphasize the fact that missing pixels will bring some spectral
contamination (i.e., aliasing) in the sampled signal. 

This becomes more evident if we calculate the spectrum of the measured
signal from equations (\ref{eq:t'(r)_miss}) and (\ref{eq:D(w)_miss}) 

\begin{equation}
T^{\prime}\left(\mathbf{w}\right)=\frac{1}{l_{1}l_{2}}\sum_{m,n=-\infty}^{\infty}E\left(\mathbf{w}_{mn}\right)T\left(\mathbf{w}-\mathbf{w}_{mn}\right).\label{eq:T'(w)_miss}\end{equation}

We see that the different replicas of $T\left(\mathbf{w}\right)$
are spaced in frequency according to equation (\ref{eq:w_mn_miss})
(compare this with the case of the ESG in equation (\ref{eq:w_mn})
where the spacing between replicas is $N$-times greater). Contrary
to the case of an ESG, a convolution with the usual weighting function
$w\left(\mathbf{r}\right)$ will not restore a low-pass filtered version
of the $t\left(\mathbf{r}\right)$ map. To make this clear, we first
define a new function $Y\left(\mathbf{w}\right)$ such that

\begin{equation}
Y\left(\mathbf{w}\right)=\sum_{m,n}E\left(\mathbf{w}_{mn}\right)T\left(\mathbf{w}-\mathbf{w}_{mn}\right).\label{eq:Y(w)_miss}\end{equation}

\noindent Then proceeding with the usual interpolation defined in
equation (\ref{eq:z_int(r)}) we find that

\begin{equation}
T_{\mathrm{int}}\left(\mathbf{w}\right)=\frac{1}{\left(l_{1}l_{2}\right)^{2}}\sum_{s,t=-\infty}^{\infty}e^{-j2\pi\mathbf{w}_{st}\cdot\mathbf{a}_{pq}}\left\{ \left[Y\left(\mathbf{w}\right)W\left(\mathbf{w}\right)\right]\otimes N\left(\mathbf{w}\right)\right\} _{\mathbf{w}=\mathbf{w}-\mathbf{w}_{st}}\label{eq:Tint(w)_miss}\end{equation}

\noindent and

\begin{equation}
t_{\mathrm{int}}\left(\mathbf{r}\right)=\left\{ \left[y\left(\mathbf{r}\right)\otimes w\left(\mathbf{r}\right)\right]\cdot n\left(\mathbf{r}\right)\right\} \cdot\frac{1}{l_{1}l_{2}}\sum_{i,k=-\infty}^{\infty}\delta\left(\mathbf{r}-\mathbf{r}_{ik}-\mathbf{a}_{pq}\right),\label{eq:tint(r)_miss}\end{equation}

\noindent where $\mathbf{a}_{pq}$ is defined in equation (\ref{eq:a_pq})
and denotes, once again, the position of the origin of the interpolated
grid. These equations show that maps resulting from the interpolation
of ISGs containing missing pixels suffer from both spectral aliasing
(from the presence of $Y\left(\mathbf{w}\right)$ in lieu of $T\left(\mathbf{w}\right)$
in eq. {[}\ref{eq:Tint(w)_miss}]) and broadening (because of the
presence of $N\left(\mathbf{w}\right)$).

We once again stress the realization that \emph{missing pixels will
bring some amount of aliasing that will be impossible to remove with
a reasonably sized weighting function. The concept of Nyquist sampling
can even lose much of its meaning and usefulness in a situation where
too many pixels are missing, or when the level of contamination due
to spectral aliasing is comparable to the noise level present in the
map.} This fact strongly underlines the necessity of performing adequate
dithers or other scanning strategies when observing with an imperfect
detector array.

We complete the analysis by performing the windowing mentioned earlier,
which transforms equation (\ref{eq:T'(w)_miss})  to

\begin{eqnarray}
t_{\mathrm{int}}^{\prime}\left(\mathbf{r}\right) & = & t_{\mathrm{int}}\left(\mathbf{r}\right)\cdot\left[\mathrm{rect}\left(\frac{x}{N_{1}l_{1}}\right)\mathrm{rect}\left(\frac{y}{N_{2}l_{2}}\right)\right]\label{eq:t'int(r)_miss}\\
T_{\mathrm{int}}^{\prime}\left(\mathbf{w}\right) & = & T_{\mathrm{int}}\left(\mathbf{w}\right)\otimes\left[N_{1}l_{1}N_{2}l_{2}\,\mathrm{sinc}\left(\pi uN_{1}l_{1}\right)\mathrm{sinc}\left(\pi vN_{2}l_{2}\right)\right].\label{eq:T'int(w)_miss}\end{eqnarray}

For reasonably large detector arrays we do not expect that the presence
of the $sinc$ functions will be of any significance due to their
spectral narrowness relative to the extent of $T\left(\mathbf{w}\right)$
(or $Y\left(\mathbf{w}\right)W\left(\mathbf{w}\right)$). Because
of this, our periodic depiction of the array, upon which our analysis
rests, is justified.

\section{Summary}

In this paper we addressed the question of astronomical image processing
from data obtained with array detectors. We defined and analyzed the
cases of evenly (ESG), regularly (RSG), and irregularly (ISG) sampled
grids for idealized and realistic detectors. We focused on the effect
of interpolation on the maps, while using a Gaussian kernel to accomplish
this task. In all cases (i.e., ESG, RSG, and ISG) we have applied
the method of weighted averages (eq. {[}\ref{eq:z_int(r)}]) to produce
a map interpolated on a finely spaced grid. 

We defined an ESG as a map where the signal to be analysed is digitized
with a simple, two-dimensional, train of Dirac distributions evenly
separated with well-defined spacings (see eq. {[}\ref{eq:diractrain}]).
Moreover, since the ESG is the simplest way to represent and analyse
a set of sampled data, we always sought to transform a non-evenly
sampled grid (i.e., a RSG or an ISG) to an ESG through the process
of interpolation. While studying the ESG, we found that the interpolation
process invariably leads to a loss in spatial resolution; this loss
grows with increasing width of the smoothing kernel (this result is
true in general, i.e., when considering RSGs and ISGs). When an ESG
is re-sampled at the same rate as the original map the interpolation
process can usually be adequately taken into account by replacing
the original signal by its convolution with the weighting function
(see eq. {[}\ref{eq:tint(r)_app}]). The same is not true in general,
however, when the final ESG is re-sampled at a rate different than
that of the original map (see eq. {[}\ref{eq:Ts(w)_b}]). This is
due to the fact that the normalization function that is intrinsic
(and necessary) to the interpolation process is not constant but a
function of position (although it is periodic with periods corresponding
to the sampling rates). The aforementioned replacement of the original
map in the interpolation process by its convolution with the weighting
function will only be adequate in such cases when the latter is sufficiently
broad relative to the spacing between samples (see Figs. \ref{fig:normalization}
and \ref{fig:ripple}). 

We defined an RSG as a generalization of an ESG such that it consists
of a combination of a number of relatively translated ESGs of similar
sampling rates. All the results obtained for the ESG can be generalized
to the RSG. 

We analyzed two different types of ISGs: the combination of relatively
translated and rotated ESGs, and ESG-like maps with missing samples
(e.g., data grids made with detectors exhibiting dead pixels). In
the first case, the interpolation process cannot be simply represented
by a convolution of the original signal with the weighting function,
as this operation must be subsequently multiplied by the normalization
function (see eq. {[}\ref{eq:tint(r)_ISG_approx}]). Because of the
irregular nature of the new map, the normalization function is not
periodic in general and will add structure to the spectrum (i.e.,
the Fourier transform) of the map. More precisely, the spectrum of
the source (filtered by the spectral profile of the weighting function)
will be broadened through its convolution with the Fourier transform
of the normalization function. This effect is a function of the size
of the weighting function, as the spatial variation of the normalization
function grows larger for smaller kernel widths. Although these results
also apply to maps exhibiting missing samples, we found that these
further suffer from spectral aliasing that may reduce or negate the
usefulness of the Nyquist sampling criterion in extreme cases. This
fact strongly underlines the necessity of performing adequate dithers
or other scanning strategies when observing with an imperfect detector
array.

\acknowledgements{M.H.'s research is funded through the NSERC Discovery Grant, Canada
Research Chair, Canada Foundation for Innovation, Ontario Innovation
Trust, and Western's Academic Development Fund programs. J.E.V. acknowledges
support from NSF grants AST 05-40882 to the California Institute of
Technology and AST 05-05124 to the University of Chicago. SHARC II
is also funded through the NSF grant AST 05-40882 to the California
Institute of Technology. SHARP is funded through the NSF grants AST
02-43156 and AST 05-05230 to Northwestern University.}

\appendix
\section{Appendix}

\label{Appendix}In this Appendix we provide a few simple derivations
to justify some of the results used in the text.

\subsection{Fourier Transform of a Dirac Train}

The Fourier transform of a Dirac train can easily be determined by
first calculating the associated Fourier series. In the one-dimensional
case we have 

\begin{equation}
\sum_{i=-\infty}^{\infty}\delta\left(x-il\right)=\frac{1}{l}\sum_{n=-\infty}^{\infty}e^{j2\pi n\frac{x}{l}},\label{eq:del(r)_app}\end{equation}

\noindent since the Fourier series for a periodic function $g\left(x\right)$
of period $l$ is defined as

\[
g\left(x\right)=\sum_{n=-\infty}^{\infty}G\left(n\right)e^{j2\pi n\frac{x}{l}},\]

\noindent with the Fourier coefficient $G\left(n\right)$ 

\[
G\left(n\right)=\frac{1}{l}\int_{-\frac{l}{2}}^{\frac{l}{2}}g\left(x\right)e^{-j2\pi n\frac{x}{l}}dx.\]

Before calculating the Fourier transform of equation (\ref{eq:del(r)_app}),
we consider the so-called duality property of the Fourier pair in
equations (\ref{eq:g(r)}) and (\ref{eq:G(w)}). More precisely, we
mean that if for a function $f\left(\mathbf{r}\right)$ we have

\[
f\left(\mathbf{r}\right)\Leftrightarrow F\left(\mathbf{w}\right),\]

\noindent then it must also be true that for $F\left(-\mathbf{r}\right)$
we have

\[
F\left(-\mathbf{r}\right)\Leftrightarrow f\left(\mathbf{w}\right),\]

\noindent as can be readily verified by inspection of equations (\ref{eq:g(r)})
and (\ref{eq:G(w)}). It follows from this that since

\[
\delta\left(\mathbf{r}-\mathbf{r}_{0}\right)\Leftrightarrow e^{-j2\pi\mathbf{w}\cdot\mathbf{r}_{0}},\]

\noindent then

\begin{equation}
e^{j2\pi\mathbf{w}_{0}\cdot\mathbf{r}}\Leftrightarrow\delta\left(\mathbf{w}-\mathbf{w}_{0}\right).\label{eq:exp_FT}\end{equation}

Using the one-dimensional version of equation (\ref{eq:exp_FT}) to
calculate the Fourier transform of equation (\ref{eq:del(r)_app})
we find that

\begin{equation}
\sum_{i=-\infty}^{\infty}\delta\left(x-il\right)\Leftrightarrow\frac{1}{l}\sum_{n=-\infty}^{\infty}\delta\left(u-\frac{n}{l}\right).\label{eq:del(u)_app}\end{equation}

The two-dimensional generalization of this result is straightforward
and leads to equations (\ref{eq:diractrain}) and (\ref{eq:dtrain_ft})
.

\subsection{The Nyquist Sampling Criterion}The \emph{Nyquist Sampling
Criterion} can be understood with equation (\ref{eq:del(u)_app})
and the product/convolution property of the Fourier transform. This
property states that the following Fourier pair is valid

\[
f\left(\mathbf{r}\right)g\left(\mathbf{r}\right)\Leftrightarrow F\left(\mathbf{w}\right)\otimes G\left(\mathbf{w}\right)\]

\noindent for two functions $f\left(\mathbf{r}\right)$ and $g\left(\mathbf{r}\right)$,
as can easily be verified from the definition of the Fourier transform.
Therefore, for the sampling of a function $t\left(x\right)$ we have

\begin{eqnarray*}
t\left(x\right)\cdot\sum_{i=-\infty}^{\infty}\delta\left(x-il\right) & \Leftrightarrow & T\left(u\right)\otimes\frac{1}{l}\sum_{n=-\infty}^{\infty}\delta\left(u-\frac{n}{l}\right)\\
 & \Leftrightarrow & \frac{1}{l}\sum_{n=-\infty}^{\infty}T\left(u-\frac{n}{l}\right),\end{eqnarray*}

\noindent which implies that the base spectrum $T\left(u\right)$
is repeated in frequency space at an interval equal to the sampling
period of $l^{-1}$. It is apparent that in order to avoid any cross-contamination
between the different spectral replicas of $T\left(u\right)$ the
following relation must be enforced

\begin{equation}
\begin{array}{cc}
T\left(u\right)=0 & \mathrm{for}\end{array}\left|u\right|\geq\frac{1}{2l}.\label{eq:Nyquist}\end{equation}

Relation (\ref{eq:Nyquist}) is the one-dimensional mathematical equivalent
of the Nyquist sampling criterion, which states that in order to recover
the base spectrum $T\left(u\right)$ from its sampled version (through
spectral filtering) the sampling frequency must be at least twice
as large as the frequency extent of $T\left(u\right)$.

\subsection{Normalization Factor}We know from our analysis that an
interpolated map $t_{\mathrm{int}}\left(\mathbf{r}\right)$ resulting
from a previously sampled data set $t^{\prime}\left(\mathbf{r}\right)$
is given by 

\[
t_{\mathrm{int}}\left(\mathbf{r}\right)=\sum_{s,t,=-\infty}^{\infty}\delta\left(\mathbf{r}-\mathbf{r}_{st}-\mathbf{a}_{pq}\right)\cdot\left\{ n\left(\mathbf{r}\right)\cdot\left[t^{\prime}\left(\mathbf{r}\right)\otimes w\left(\mathbf{r}\right)\right]\right\} ,\]

\noindent which we transform slightly to

\begin{equation}
t_{\mathrm{int}}\left(\mathbf{r}\right)=\left[n\left(\mathbf{r}\right)\cdot\sum_{s,t=-\infty}^{\infty}\delta\left(\mathbf{r}-\mathbf{r}_{st}-\mathbf{a}_{pq}\right)\right]\cdot\left[t^{\prime}\left(\mathbf{r}\right)\otimes w\left(\mathbf{r}\right)\right].\label{eq:tint(r)_appendix}\end{equation}

As usual $n\left(\mathbf{r}\right)$ and $w\left(\mathbf{r}\right)$
are the normalization and weighting functions, respectively, and the
vectors $\mathbf{r}_{st}$ and $\mathbf{a}_{pq}$ are given by equations
(\ref{eq:r_ik}) and (\ref{eq:a_pq}). When the original map is an
ESG the normalization function is periodic and can therefore be expanded
with a two-dimensional Fourier series

\begin{equation}
n\left(\mathbf{r}\right)=\sum_{i,k=-\infty}^{\infty}N\left(i,k\right)e^{j2\pi\mathbf{w}_{ik}\cdot\mathbf{r}}\label{eq:n(r)_series}\end{equation}

\noindent with $N\left(i,k\right)$ the corresponding Fourier coefficient
and $\mathbf{w}_{ik}$ given by equation (\ref{eq:w_mn}).

From equations (\ref{eq:exp_FT}) and (\ref{eq:n(r)_series}) we can
write the Fourier transform of equation (\ref{eq:tint(r)_app}) as

\begin{eqnarray*}
T_{\mathrm{int}}\left(\mathbf{w}\right) & = & \left\{ \left[\sum_{i,k=-\infty}^{\infty}N\left(i,k\right)\delta\left(\mathbf{w}-\mathbf{w}_{ik}\right)\right]\otimes\left[\frac{1}{l_{1}l_{2}}\sum_{m,n=-\infty}^{\infty}\delta\left(\mathbf{w}-\mathbf{w}_{mn}\right)e^{-j2\pi\mathbf{w}\cdot\mathbf{a}_{pq}}\right]\right\} \\
 &  & \,\,\,\otimes\left[T^{\prime}\left(\mathbf{w}\right)W\left(\mathbf{w}\right)\right]\\
 & = & \left[\frac{1}{l_{1}l_{2}}\sum_{i,k=-\infty}^{\infty}N\left(i,k\right)\sum_{m,n=-\infty}^{\infty}\delta\left(\mathbf{w}-\mathbf{w}_{mn}-\mathbf{w}_{ik}\right)e^{-j2\pi\left(\mathbf{w}-\mathbf{w}_{ik}\right)\cdot\mathbf{a}_{pq}}\right]\\
 &  & \,\,\,\otimes\left[T^{\prime}\left(\mathbf{w}\right)W\left(\mathbf{w}\right)\right]\\
 & = & \left[\frac{1}{l_{1}l_{2}}\sum_{i,k=-\infty}^{\infty}N\left(i,k\right)e^{-j2\pi\mathbf{w}_{ik}\cdot\mathbf{a}_{pq}}\sum_{m,n=-\infty}^{\infty}\delta\left(\mathbf{w}-\mathbf{w}_{mn}-\mathbf{w}_{ik}\right)e^{-j2\pi\mathbf{w}\cdot\mathbf{a}_{pq}}\right]\\
 &  & \,\,\,\otimes\left[T^{\prime}\left(\mathbf{w}\right)W\left(\mathbf{w}\right)\right],\end{eqnarray*}

\noindent but since

\[
\sum_{m,n=-\infty}^{\infty}\delta\left(\mathbf{w}-\mathbf{w}_{mn}-\mathbf{w}_{ik}\right)=\sum_{m^{\prime},n^{\prime}=-\infty}^{\infty}\delta\left(\mathbf{w}-\mathbf{w}_{m^{\prime}n^{\prime}}\right)\]

\noindent with $m^{\prime}=m+i$ and $n^{\prime}=n+k$, then

\begin{eqnarray*}
T_{\mathrm{int}}\left(\mathbf{w}\right) & = & \left\{ \left[\sum_{i,k=-\infty}^{\infty}N\left(i,k\right)e^{-j2\pi\mathbf{w}_{ik}\cdot\mathbf{a}_{pq}}\right]\left[\frac{1}{l_{1}l_{2}}\sum_{m^{\prime},n^{\prime}=-\infty}^{\infty}\delta\left(\mathbf{w}-\mathbf{w}_{m^{\prime}n^{\prime}}\right)e^{-j2\pi\mathbf{w}\cdot\mathbf{a}_{pq}}\right]\right\} \\
 &  & \,\,\,\otimes\left[T^{\prime}\left(\mathbf{w}\right)W\left(\mathbf{w}\right)\right].\end{eqnarray*}

However, we can use equation (\ref{eq:n(r)_series}) one more time
to transform this last relation to

\[
T_{\mathrm{int}}\left(\mathbf{w}\right)=n\left(\mathbf{a}_{pq}\right)\left[\frac{1}{l_{1}l_{2}}\sum_{m,n=-\infty}^{\infty}\delta\left(\mathbf{w}-\mathbf{w}_{mn}\right)e^{-j2\pi\mathbf{w}\cdot\mathbf{a}_{pq}}\right]\otimes\left[T^{\prime}\left(\mathbf{w}\right)W\left(\mathbf{w}\right)\right],\]

\noindent and

\begin{equation}
t_{\mathrm{int}}\left(\mathbf{r}\right)=n\left(\mathbf{a}_{pq}\right)\sum_{s,t=-\infty}^{\infty}\delta\left(\mathbf{r}-\mathbf{r}_{st}-\mathbf{a}_{pq}\right)\cdot\left[t^{\prime}\left(\mathbf{r}\right)\otimes w\left(\mathbf{r}\right)\right].\label{eq:tint(r)_const_app}\end{equation}

Evidently $n\left(\mathbf{a}_{pq}\right)$ is constant for a given
interpolated map, but will vary as a function of the displacement
of the new sampling grid (defined by $\mathbf{a}_{pq}$) relative
to the original one. Equation (\ref{eq:tint(r)_const_app}) leads
to (and justifies) equation (\ref{eq:tint(r)_app}) provided that
we set $c=n\left(\mathbf{a}_{pq}\right)$.

\subsection{Fourier Transform of a Rotated Map}Finally, we prove
the result used in \S \ref{sec:ISG} that the Fourier transform of
a rotated map is the rotated Fourier transform of the unrotated map.
To do so we subject a two-dimensional map $g\left(\mathbf{r}\right)$
to a rotation $\mathbf{R}$ and calculate the Fourier transform of
the transformed map $g\left(\mathbf{Rr}\right)$ with

\[
G^{\prime}\left(\mathbf{w}\right)=\int_{-\infty}^{\infty}g\left(\mathbf{Rr}\right)e^{-j2\pi\mathbf{w}\cdot\mathbf{r}}dxdy.\]

We now make the change of variable $\mathbf{r}^{\prime}=\mathbf{Rr}$
to get 

\begin{eqnarray}
G^{\prime}\left(\mathbf{w}\right) & = & \int_{-\infty}^{\infty}g\left(\mathbf{r}^{\prime}\right)e^{-j2\pi\mathbf{w}\cdot\left(\mathbf{R}^{-1}\mathbf{r}^{\prime}\right)}dx^{\prime}dy^{\prime}\nonumber \\
 & = & \int_{-\infty}^{\infty}g\left(\mathbf{r}^{\prime}\right)e^{-j2\pi\left(\mathbf{R}\mathbf{w}\right)\cdot\mathbf{r}^{\prime}}dx^{\prime}dy^{\prime},\label{eq:G'(w)_app}\end{eqnarray}

\noindent where the last transformation was made possible by the fact
that the inverse of a rotation matrix equals its transpose. We therefore
find from equation (\ref{eq:G'(w)_app}) that if

\[
g\left(\mathbf{r}\right)\Leftrightarrow G\left(\mathbf{w}\right),\]

\noindent then

\[
g\left(\mathbf{Rr}\right)\Leftrightarrow G\left(\mathbf{Rw}\right).\]

\clearpage

\clearpage

\clearpage

\clearpage

\clearpage

\clearpage

\end{document}